 \documentclass[prl,twocolumn,floatfix,superscriptaddress]{revtex4}
\usepackage{dcolumn,amsmath}
\usepackage[dvips]{graphicx}
\usepackage{bm}
\usepackage{hyperref}   %%% tav added for \cite{} following for .pdf

\def\veps{\varepsilon}

\newcommand{\eref}[1]{(\ref{#1})}
\newcommand{\Eref}[1]{Eq.~(\ref{#1})}
\newcommand{\tref}[1]{Table~\ref{#1}}

\begin{document}
%############################################################
%\large
%???
%\title{Towards the electron EDM search: Theoretical study of HfF$^+$}
 \title{Theoretical study of
%ap 
%low-lying electronic terms and transition moments for
%end ap 
HfF$^+$ for the electron EDM search}

\author{A.N.\ Petrov}\email{anpetrov@pnpi.spb.ru}
\altaffiliation [Also at ] {St.-Petersburg State University, St.-Petersburg,
        Russia}
\author{N.S.\ Mosyagin}
%\author{T.A.\ Isaev}
\author{A.V.\ Titov} \altaffiliation [Also at ]
{St.-Petersburg State University, St.-Petersburg, Russia}
\affiliation
%\address
{Petersburg Nuclear Physics Institute, Gatchina,
             Leningrad district 188300, Russia}

%\date{\today}

\begin{abstract}
 We report {\it ab initio} relativistic correlation calculations of potential
 curves and electric dipole transition moments for ten low-lying electronic
 states, effective electric field on the electron, hyperfine constants and
 radiative lifetimes for the $^3\Delta_1$ state of cation of the heavy
 transition metal fluoride HfF$^+$, which it is suggested to be used in
 experiments to search for the electric dipole moment of the electron. It is
 obtained that HfF$^+$ has deeply bound $^1\Sigma^+$ ground state; its
 dissociation energy is $D_e=6.4$ eV.
%  $D_e=51~685~{\rm cm}^{-1}$.
 The $^3\Delta_1$ state is obtained as the relatively long-lived,
with lifetime equal to about 0.4 s.,
first excited state lying about 0.2 eV 
%  1633 cm$^{-1}$
 higher. The calculated effective electric field, $E_{\rm eff}=W_d|\Omega|$,
 acting on an electron in this state is $5.84\times10^{24}{\rm Hz}/e~{\rm
 cm}$. The obtained hyperfine constants are $A_{\parallel}= -1239$~MHz for the
 $^{177}$Hf nucleus and $A_{\parallel}= -58.1$~MHz for the $^{19}$F nucleus.
\end{abstract}

\maketitle

%=========================================================================
\section{Introduction.}

 The search for the electric dipole moment (EDM) of the electron ($d_e$ or
 eEDM below) remains one of the most fundamental problems in physics.
% in particular, in connection with recent discovery of the neutrino
% oscillation, which influence seriously on the modern theory of nature of
% electroweak interaction.
 Up to now only upper limits for $|d_e|$ were obtained. The tightest bound on
 $d_e$ was given in the experiment on the atomic Tl beam \cite{Regan:02},
 which established an upper bound of $|d_e|<1.6\times10^{-27}\ e~\rm{cm}$
 ($e$ is the charge of the electron).  Molecular systems provide a way to get
 much enhanced sensitivity, since the effective intramolecular electric field
 acting on electrons in polar molecules can be five or more orders of
 magnitude higher than the strongest field available in the laboratory
 \cite{Sushkov:78, Gorshkov:79, Titov:06amin}.

 The new generation of the eEDM experiments, employing polar heavy-atom
 molecules, is expected to reach sensitivity of $10^{-30}{-}10^{-28} e~{\rm
 cm} /\sqrt{\textrm{day}}$ (e.g., see \cite{Sauer:06a}).  Their results are
 expected to dramatically influence all the popular extensions of the
 Standard model, in particular supersymmetry, even if bounds on the $P,T$-odd
 effects compatible with zero are obtained (see \cite{Ginges:04, Erler:05}
 and references therein). These studies include the beam experiments carried
 out on YbF molecular radicals by Hinds and co-workers \cite{Sauer:06a}, and
 the vapor cell experiment on the metastable $a(1)$ state of PbO prepared by
 the group of DeMille and co-workers    (see \cite{Kawall:04b} and references
 therein). New ways of searching for the eEDM, using trapped cold molecular
 cations, were investigated during the last years by Cornell and co-workers.
 The first candidate was HI$^+$ \cite{Stutz:04}, but subsequent estimate
 \cite{Ravaine:05a} and accurate calculation \cite{Isaev:05a} have shown that
 this cation has rather small $E_{\rm eff}$ (see below).
%  EDM enhancement factor, $d_{\rm mol}/d_e$, (where $d_{\rm mol}$ is the
%  P,T-odd EDM of a molecule induced by the eEDM, see \cite{Flambaum:76,
%  Sushkov:78, Ginges:04, Titov:06amin} for details).
 Then some
% better motivated
 other candidates were considered, HfH$^+$ \cite{Sinclair:05Aa}, HfF$^+$
 \cite{Cornell:06PC} etc., having $^3\Delta_1$ as the ground or probably
 long-lived excited state.  Even experimental study of spectroscopic
 properties for these cations is a very difficult problem, and there are no
 such data measured up to date though they are already required to analyze
 basic stages of the eEDM experiment \cite{Cornell:06PC}. In turn, the modern
 relativistic computational methods can now give reliable answers to almost
 all the questions of interest even for compounds of heavy transition metals
 as is in our case.  In the paper \cite{Petrov:07a} a theoretical study of
 the required spectroscopic properties, $E_{\rm eff}$ and hyperfine
 structure, of HfF$^+$ as a candidate for the eEDM experiment was performed.
 In particular, it was shown that $^3\Delta_1$ is not the ground state, its
 radiative lifetime was estimated as 0.5 sec.  In the present paper the above
 study is disscused in more details and, besides, new obtained data are
 given. In particular, the more accurate data for the transition dipole
 moment $^1\Sigma^+ \rightarrow {^3}\Delta_1$ and, consequently, more
 reliable values for the radiative lifetimes of the lowest vibrational levels
 of the $^3\Delta_1$ are presented.  Also the obtained data for other
 transition dipole moments can be used to study possible schemes of
 populating the $^3\Delta_1$ state.  With using the Dirac-Hartree-Fock
 calculations it was suggested in \cite{Petrov:07a} that influence of $4f$
 electrons of Hf is negligible for the excitation energy $^1\Sigma^+
 \rightarrow {^3}\Delta_1$. Taking into account the importance of this
 conlusion for the accuracy analysis, in the present paper the influence of
 correlation of the $4f$ electrons is studied here.

%================================================================
\section{$E_{\rm eff}$ and hyperfine structure.}

 One of the most important features of such experiments is that knowledge of
 the effective electric field $E_{\rm eff}$ seen by an unpaired electron is
 required for extracting $d_e$ from the measurements. $E_{\rm eff}$ cannot be
 obtained in an experiment; rather, electronic structure calculations are
 required for its evaluation.  It is given by the expression $E_{\rm
 eff}=W_d|\Omega|$, where $W_d$ is a parameter of the P,T-odd molecular
 Hamiltonian that is presented in Refs.\, \cite{Kozlov:87, Kozlov:95,
 Titov:06amin},

\begin{equation}
   W_d = \frac{1}{\Omega d_e}
   \langle \Psi|\sum_iH_d(i)|\Psi
 \rangle,
\end{equation}
 where $\Psi$ is the wave function for the $^3\Delta_1$ state, and $\Omega=
 \langle\Psi|\bm{J}\cdot\bm{n}|\Psi\rangle = \pm1 $, $\bm{J}$ is the total
 electronic momentum, $\bm{n}$ is the unit vector along the molecular axis
 directed from Hf to F,

\begin{eqnarray}
    H_d=2d_e
    \left(\begin{array}{cc}
    0 & 0 \\
    0 & \bm{\sigma E} \\
    \end{array}\right)\ ,
\label{Wd}
\end{eqnarray}
 $\bm{E}$
%  Better: $\bm{E} = ...$
 is the inner molecular electric field, and $\bm{\sigma}$ are the Pauli
 matrices.  The $E_{\rm eff}$ value for the $^3\Delta_1$ state was estimated
 in the scalar-relativistic approximation by Meyer {\it et al.} (see the note
 added at the end of \cite{Meyer:06a})
% они должны различать состояния $^3\Delta_1$, $^3\Delta_2$ и $^3\Delta_3$
% так как у них (берем ML=+2) MS=-1, MS=0 и MS=+1 для первого и последнего
% E_eff будет отличаться знаком (насчет знака могли в частности здесь
% ошибиться), а для $^3\Delta_2$ (MS=0), (в скалярно релятивистском
% приближении) $W_d=0$ 
 and the method used for calculation of $E_{\rm eff}$ is close in essence to
 that developed by Titov earlier \cite{Titov:85Dis} and applied to two-step
 calculations of the PbF molecule \cite{Titov:85Dis, Kozlov:87}.  In paper
 \cite{Petrov:07a} a more reliable value of $E_{\rm eff}$ is calculated using
 the advanced two-step technique developed by Titov and co-workers later
 \cite{Titov:96, Mosyagin:98, Petrov:02, Petrov:05a}.
%
%  The eEDM enhancement factor, $d_{\rm mol}/d_e = E_{\rm eff}/E_{\rm ext}$
%  (where $E_{\rm ext}$ is some external field required to polarize the
%  molecule, see \tref{core}), can reach $10^8$ (???) since $E_{\rm ext}$ is
%  only of order of 10~eV (???) magnitude for such systems as HfF$^+$ in the
%  $^3\Delta_1$ state \cite{Cornell:06PC}.
%
 The hyperfine constants for $^{177}$Hf and $^{19}$F nuclei
 ($A_{\parallel}[\rm{Hf}]$ and $A_{\parallel}[\rm{F}]$) for the $^3\Delta_1$
 state of HfF$^+$ are also calculated.  The hyperfine constants are
 determined by expression \cite{Dmitriev:92}

 \begin{eqnarray}
   A_{\parallel}[\rm{Hf}(\rm{F})]=\frac{1}{\Omega} \frac{\mu_{\rm Hf(F)}}{I_{\rm Hf(F)}}
   \langle
   \Psi|\sum_i(\frac{\bm{\alpha}_i\times \bm{r}_i}{r_i^3})
_Z|\Psi
   \rangle~,
 \label{All}
\end{eqnarray}
 where $I_{\rm Hf(F)}$ and $\mu_{\rm Hf(F)}$ are the spin momentum and
 magnetic moment of $^{177}$Hf($^{19}$F), $\bm{\alpha}_i$ are the Dirac
 matrices for the $i$-th electron, and $\bm{r}_i$ is its radius-vector in the
 molecular coordinate system centered on the Hf or F atom.  When the
 experimental value $A_{\parallel}[\rm{Hf}]$ is measured, it provides an
 accuracy check for the calculated $E_{\rm eff}$ value. Both
 $A_{\parallel}[\rm{Hf}]$ and $A_{\parallel}[\rm{F}]$ values are useful for
 identifying HfF$^+$ by its spectrum from other species in the experiment.
%
% Moreover, preparation of the HfF$^+$ cations in the $^3\Delta_1$ state for
% the eEDM experiment, registration of eEDM signals etc.\ requires knowledge
% of the spectroscopic properties. In particular, a question to be solved was
% which state was the ground state. If $^3\Delta_1$ is not the ground state,
% its radiative lifetime is also required.  In the paper \cite{Petrov:07a}
% these data were obtained , though precise study of the compounds with
% transition elements is a difficult problem for modern molecular theory.
% The needed spectroscopic information cannot be obtained now from other
% sources. 

%====================================================================
\section{Methods and calculations}

% The generalized relativistic effective core potential (GRECP)
% \cite{Titov:00a} simulating interaction of valence electrons  (12 electrons
% of Hf) with the explicitly excluded $1s$ to $4f$ electrons of Hf (60 core
% electrons) is used in 20-electron calculations of HfF$^+$.  In ten-electron
% calculations, which are substantially less time consuming, $5s$ and $5p$
% spinors of hafnium and $1s$ orbital of fluorine are frozen from the states
% averaged over the nonrelativistic configurations $5d^2 6s^{0.6} 6p^{0.4}$
% for Hf$^+$ and $2s^2 2p^5$ for F, and not treated explicitly. 

%=====================================================================
 \subsection{The GRECP method.}
 \label{sGRECP}

 When core electrons of a heavy-atom molecule do not play an active role
  (i.e., their relaxation in the molecule is negligible)
 the effective Hamiltonian with relativistic effective core potential
 (RECP) can be presented in the form

\begin{equation}
   {\bf H}^{\rm Ef}\ =\ \sum_{i_v} \Bigl[{\bf h}^{\rm Schr}(i_v) +
          {\bf U}^{\rm Ef}(i_v)\Bigr] + \sum_{i_v > j_v} \frac{1}{r_{i_v j_v}}\ .
 \label{Heff2}
\end{equation}
 Hamiltonian~\eref{Heff2} is written only for a valence subspace of electrons,
 which are treated explicitly and denoted by indices $i_v$ and $j_v$. In
 practice, this subspace is often extended by inclusion of some outer core
 shells for better accuracy. In \Eref{Heff2}, ${\bf h}^{\rm Schr}$
 is the one-electron Schr\"odinger Hamiltonian

\begin{equation}
     {\bf h}^{\rm Schr}\ = - \frac{1}{2} {\vec \nabla}^2 
     - \frac{Z_{ic}}{r}\ ,
 \label{Schr}
\end{equation}
 where $Z_{ic}$ is the charge of the nucleus decreased by the number of inner
 core electrons. ${\bf U}^{\rm Ef}$ is an RECP (or relativistic
 pseudopotential (PP)) operator that can be written in the separable (e.g.,
 see Ref.~\cite{Theurich:01} and references therein) or radially-local
 (semi-local)\cite{Ermler:88} approximations when the valence pseudospinors
 are smoothed in heavy-atom cores. This smoothing allows one to reduce the
 number of primitive Gaussian basis functions required for appropriate
 description of valence spinors in subsequent molecular calculations and also
 to exclude the small components of the four-component Dirac spinors from the
 RECP calculations, with relativistic effects being taken into account by
 $j$-dependent effective potentials. Contrary to the four-component wave
 function used in Dirac-Coulomb(-Breit) calculations, the pseudo-wave
 function in the RECP case can be both two- and one-component.  

 Besides, the generalized RECP (GRECP) operator\cite{Titov:99, Titov:00a} can
 be used in \Eref{Heff2} that includes the radially-local, separable and
 Huzinaga-type\cite{Bonifacic:74} relativistic PPs as its components and some
 special cases. The GRECP concept was introduced and developed in a series of
 papers (see Refs.~\cite{Titov:99, Titov:00a, Petrov:04b, Mosyagin:05a,
 Mosyagin:05b, Titov:05b} and references therein). In contrast to other RECP
 methods, GRECP employs the idea of separating the space around a heavy atom
 into three regions: inner core, outer core and valence, which are treated
 differently. It allows one to attain theoretically any desired accuracy,
 while requiring moderate computational efforts since the overall accuracy is
 limited in practice by possibilities of correlation methods.

 Two of the major features of the GRECP version with the separable correction
 described here are generating of the effective potential components for the
 pseudospinors which may have nodes, and addition of non-local separable
 terms with projectors on the outer core pseudospinors to the conventional
 semi-local RECP operator. The problem of division by zero appearing in the
 cases of pseudospinors with nodes is overcome in the GRECP method by
 interpolating the corresponding potentials in the vicinity of these
 nodes\cite{Titov:91, Mosyagin:94}. It was shown both theoretically and
 computationally that the interpolation errors are small enough. That allows
 us to generate different potentials for the cases of outer core and valence
 pseudospinors with the same quantum numbers $l$ and $j$, unlike the
 conventional RECP approach. In turn, the non-local separable terms in the
 GRECP operator account for difference between these potentials, which in the
 outer region is defined by smoothing within the inner core as is shown in
 Ref.~\cite{Titov:03Ae, Titov:02Dis, Titov:05e} and in many cases this
 difference can not be neglected for ``chemical accuracy'' (about 1~kcal/mol
 or 350~cm$^{-1}$) of valence energies. The more circumstantial description
 of distinctive features of the GRECP as compared to the original RECP
 schemes is given in Refs.~\cite{Mosyagin:98c, Titov:00b}. Some other GRECP
 versions are described and discussed in details in Refs.~\cite{Titov:99,
 Titov:00a, Mosyagin:05b}.

 The GRECP operator in the spinor representation\cite{Tupitsyn:95, Titov:99}
 is naturally used in atomic calculations. The spin-orbit representation of
 this operator which can be found in Ref.~\cite{Mosyagin:97, Titov:99} is
 more efficient in practice being applied to calculation of molecules.
 Despite the complexity of expression for the GRECP operator, the calculation
 of its one-electron integrals is not notably more expensive than that for
 the case of the conventional radially-local RECP operator.

% The (G)RECP operator simulates, in particular, interactions of the explicitly 
% treated electrons with those which are excluded from the (G)RECP calculations. 
% General justification of the possibility to simulate the Breit effects by 
% means of an one-electron (G)RECP operator with good accuracy and the scheme 
% of such (G)RECP generation are presented in 
% Refs.~\cite{Petrov:04b, Mosyagin:05a}. The conventional Coulomb operator 
% for two-electron interactions and the point nuclear model should be used 
% in calculations with such (G)RECPs. However, they will account for the Fermi 
% nuclear charge model that is close to the experimental distribution.
% Moreover, the Breit interactions of the electrons from the state used for the
% (G)RECP generation with the electrons explicitly treated in the subsequent
% calculations will be simulated by the (G)RECP (in some sense, the Breit 
% interaction is ``frozen'' here).

 The GRECP simulating interaction of 12 outercore and valence electrons of Hf 
 with the explicitly excluded $1s$ to $4f$ electrons (60 inner core
 electrons) is used in 20-electron calculations of HfF$^+$.

%=====================================================================
 \subsection{Freezing the innermost shells from the outer core space.}
 \label{sOC-Fr}

 The ``freezing'' of innermost shells from the outer core space of electrons
 within the ``small core'' GRECPs is sometimes required because the accuracy
 of the GRECPs generated directly for a given number of explicitly treated
 electrons cannot always correspond to the accuracy of the conventional
 ``frozen core'' approximation with the same space of explicitly treated
 electrons (without accounting for the frozen states). That space is usually
 chosen as a minimal one required for attaining a given accuracy. It was
 noted in Refs.~\cite{Mosyagin:94, Tupitsyn:95} that using essentially
 different smoothing radii for spinors with different $lj$ is not expedient
 since the (G)RECPs errors are mainly accumulated by the outermost from them.
 In turn, explicit treatment of all of the outer core shells of an atom with
 the same principal quantum number is not usually reasonable in molecular
 (G)RECP calculations because of essential increase in computational efforts
 without serious improvement of accuracy. A natural way out is to ``freeze''
 the innermost of them before performing molecular calculation but this can
 not be done directly if the spin-orbit molecular basis set is used whereas
 the core shells should be better frozen as spinors.

 In order to exclude (``freeze'') explicitly those innermost shells (denoted by
 indices $f$ below) from molecular (G)RECP calculation without changing the
 radial node structure of other (outermore core and valence) shells in the core
 region, the energy level shift technique can be applied to overcome the above
 contradiction \cite{Titov:99, Titov:01}.  Following Huzinaga {\it et
 al.} \cite{Bonifacic:74}, one should add the effective core operator 
 ${\bf U}^{\rm Ef}_{\rm Huz}$ containing the Hartree-Fock field operators, the
 Coulomb ($\tilde {\bf J}$) and spin-dependent exchange ($\tilde {\bf K}$)
 terms, over these core spinors together with the level shift terms to the
 one-electron part of the Hamiltonian~\eref{Heff2}:

\begin{equation}
  {\bf U}^{\rm Ef}_{\rm Huz} =
    \bigl({\bf \tilde J{-}\tilde K}\bigr)[\tilde f_{n_flj}]\ +
      \sum_{n_f,l,j} B_{n_flj}\
       \bigl|\tilde f_{n_flj} \bigr\rangle \bigl\langle \tilde f_{n_flj}\bigr|
 \label{OC_Fr-1}
\end{equation}
\[
          \quad (\mbox{i.e.}\ \ \varepsilon_{n_flj} \to
                 \varepsilon_{n_flj}{+}B_{n_flj})\ ,
\]		 
 where $|\tilde f_{n_flj} \rangle\langle \tilde f_{n_flj}|$ are the
 projectors on the core spinors $\tilde f_{n_flj}$ and $\varepsilon_{n_flj}$
 are their one-electron energies.  The $B_{n_flj}$ parameters are presented
 as $M|\varepsilon_{n_flj}|$ in our codes and $M>1$ to prevent occupying the
 corresponding states in calculations; it is usually selected as $M \gg 1$ in
 our studies.  Such nonlocal terms are needed in order to prevent collapse of
 the valence electrons to the frozen core states. They introduce some ``soft
 orthogonality constraint'' between the ``frozen'' and other explicintly
 treated outermore core and valence electronic states.

 All the terms with the frozen core spinors (the level shift operator
 and exchange interactions) can be transformed to the spin-orbit
 representation in addition to the spin-independent Coulomb term,
 using the identities for the ${\bf P}_{lj}$ projectors\cite{Hafner:79}:

\begin{equation}
        {\bf  P}_{l,j=l\pm 1/2}\
         =\ \frac{1}{2l{+}1} \Biggl[ \biggl(l +
            \frac{1}{2} \pm \frac{1}{2}\biggr)
            {\bf P}_l \pm
          2 {\bf P}_l\
          \vec{\bf l}{\cdot}\vec{\bf s}\ {\bf P}_l \Biggr]\ ,
\label{Oper_Pnljs}
\end{equation}
\[
  {\bf P}_{lj} = \sum\limits_{m_j=-j}^j
    \bigl| ljm_j \bigl\rangle \bigr\langle ljm_j \bigr|\ ,
\]    
\[
  {\bf P}_{l} =
     \sum\limits_{m_l=-l}^l \bigl| lm_l \bigl\rangle \bigr\langle lm_l \bigr|\ .
\]     
 where $\vec{\bf l}$ and $\vec{\bf s}$ are operators of the orbital and spin
 momenta, $| ljm_j \rangle\langle ljm_j |$ is the projector on the
 two-component spin-angular function $\chi_{ljm_j}$, $| lm_l \rangle\langle
 lm_l |$ is the projector on the spherical function $Y_{lm_l}$.

 More importantly, these outer core pseudo{\it spinors} can be frozen in
 calculations with the {\it spin-orbit} basis sets and they can already be
 frozen at the stage of calculation of the one-electron matrix elements of
 the Hamiltonian, as implemented in the {\sc molgep} code \cite{MOLGEP}. Thus,
 any integrals with indices of the frozen spinors are completely excluded
 after the integral calculation step.  The multiplier $M{=}30$ was chosen in
 the present molecular calculations to prevent mixing the shifted core states
 to the wavefunction due to correlations but not to get poor reference
 wavefunction in the initial spin-averaged calculations at the same time (as
 would be for $M\to\infty$).

 In fact, the {\it combined} GRECP version, with separable and Huzinaga-type
 terms, is a new pseudopotential treating some minimal number of electrons 
 explicitly but
 which already provide the accuracy approaching to that of the frozen core
 approximation.  The efficiency of using the ``freezing'' procedure within
 the GRECP method was first studied in calculations of Tl \cite{Titov:99}
 and TlH \cite{Titov:01}.

% The freezing technique discussed above can be efficiently applied to those
% outer core shells for which the spin-orbit interaction is clearly more
% important than the correlation and relaxation effects. If the latter effects
% are neglected entirely or taken into account within ``correlated'' GRECP
% versions\cite{Mosyagin:05b}, the corresponding outer core pseudospinors 
% can be frozen and the
% spin-orbit basis sets can be successfully used for other explicitly treated
% shells. 
% This is true for the $6p_{1/2,3/2}$ subshells in E112, contrary to the
% case of the $6d_{3/2,5/2}$ subshells. Freezing the outer core pseudospinors
% allows one to optimize an atomic basis set only for the orbitals which are
% varied or explicitly correlated in subsequent calculations, thus avoiding the
% basis set optimization for the frozen states and reducing the number of the
% calculated and stored two-electron integrals.  Otherwise, if the $6p$ shell
% should be correlated explicitly, a spinor basis set can be more appropriate
% than the spin-orbit one in a molecular calculation.

 In ten-electron calculations, which are substantially less time consuming,
 $5s$ and $5p$ spinors of hafnium and $1s$ orbital of fluorine are frozen
 from the states averaged over the nonrelativistic configurations $5d^2
 6s^{0.6} 6p^{0.4}$ for Hf$^+$ and $2s^2 2p^5$ for F, and not treated
 explicitly.

%=====================================================================
 \subsection{The SODCI method.}
 \label{sSODCI}

 The spin-orbit direct configuration interaction (SODCI) method is well
 described in papers~\cite{Buenker:99,Alekseyev:04a}. In the current version
 of the method, calculations are carried out in the $\Lambda S$ basis set of
 many-electron spin-adapted (and space symmetry-adapted) functions (SAFs).
 The different $\Lambda S$ sets of SAFs are collected together in accord to
 the relativistic double-group symmetry requirements for the final
 configuration interaction (CI) calculation.  In the present study of the
 molecule having the $^{2S'+1}\Lambda'$ leading term, configurations from all
 the symmetry allowed $\Lambda S$ sets with
% $|S-S'|\leq 2$
 $S \leq 2$ are included into calculations.

 All the possible singly and doubly excited configurations with respect to
 some reference configurations are generated (see below for choice of the
 reference configurations). A generated configuration is included in the
 final CI space if its addition to the reference set leads to lowering in the
 total energy by the value more than some threshold $T$. This lowering is
 estimated with the help of the $A_k$ version of perturbation
 theory~\cite{Gershgorn:68}, in which the correlation and spin-orbit
 interaction are considered as perturbations and the wavefunction (obtained
 from the CI calculation in the space of the reference configurations for all
 the $\Lambda S$ irreducible representations) is taken as a zero
 approximation (see~\cite{Titov:01, Petrov:05a} for details). The lowerings
 in the total energies for the unselected configurations are employed for the
 $T{=}0$ threshold extrapolation. The generalized multireference
 analogue~\cite{Bruna:80} of the Davidson correction~\cite{Davidson:74} (full
 CI correction) is also calculated.

 In the present calculations, those configurations are chosen as the
 reference (main) configurations which give the largest contribution to the
 wavefunction (i.e.\ have the largest square of the absolute value of the CI
 coefficient ($C_I$), and thus, that $C_{ref}^2 \equiv  \sum_{I \in {\rm
 ref}} |C_I|^2 = C$, where $C=0.973$ for ten-electron and $C=0.942$ for
 20-electron calculations.  The configurations are obtained from results of
 the preliminary SODCI calculations in the relativistic double-group symmetry
 with the large threshold $T$.  New SODCI calculation is then carried out
 with the smaller thresholds.  The $C$ is taken the same for each point on
 the potential curve \cite{Mosyagin:02}.  Davidson and other corrections
 estimating contributions for higher than double excitations have an
 essential dependence on
%!!! here $C_{ref}^2$ .ne $C$!
 $C_{ref}^2$. The above selection criterion allows one to stabilize these
 corrections for different internuclear distances. This is important because
 the reliability of those corrections has a significant dependence on these
 values.

%=====================================================================
 \subsection{Basis sets and property calculations.}

 The generalized
 correlation atomic basis set \cite{Mosyagin:00,Isaev:00}
 ($12s16p16d10f10g$)/[$6s5p5d3f1g$] is constructed for Hf. The ANO-L
 ($14s9p4d3f$)/[$4s3p2d1f$] atomic basis set listed in the {\sc molcas~4.1}
 library \cite{MOLCAS} was used for fluorine.  The molecular orbitals are
 obtained by the complete active space self-consistent field (CASSCF) method
 \cite{Olsen:88,MOLCAS} with the spin-averaged part of the GRECP
 \cite{Titov:99}, i.e.\ only scalar-relativistic effects are taken into
 account at this stage.  In the CASSCF method, orbitals are subdivided into
 three groups: inactive, active, and virtual. Inactive orbitals are doubly
 occupied in all the configurations; all possible occupations are allowed for
 active orbitals, whereas virtual orbitals are not occupied. So the wave
 function is constructed as a full configuration-interaction expansion in the
 space of active orbitals, and both active and inactive orbitals are optimized
 for subsequent correlation calculations of HfF$^+$. According to the $C_{2v}$ point
 group classification scheme used in our codes, five orbitals in A$_1$, four in B$_1$ and B$_2$,
 and two in A$_2$ irreducible representations (irreps) are included in the active space.
 In ten-electron calculations, one orbital in the A$_1$ irrep (which is mainly
 the $2s$ orbital
 of F) belongs to the inactive space. In 20-electron CASSCF calculations, the
 $5s$ and $5p$ orbitals of Hf and $1s$ orbital of F are added to the space of
 inactive orbitals.

% Next, the spin-orbit direct configuration interaction (SODCI) approach
% \cite{Buenker:74, Buenker:99, Alekseyev:04a} (modified by us to account
% for spin-orbit interaction in configuration selection procedure
% \cite{Titov:01}) with selected single and double excitations from some
% multiconfigurational reference states is employed on the sets of different
% {$\Lambda$}S many-electron spin- and space-symmetry adapted basis functions.
% Details of the features of constructing the reference space and the
% selection procedure are given in Refs.~\cite{Titov:01, Mosyagin:02, Petrov:05a}.

 The ten lowest states with the leading configurations
 $[...]\sigma_1^2\sigma_2^2$ ($^1\Sigma^+$),
 $[...]\sigma_1^2\sigma_2^1\delta^1$ ($^3\Delta_{1,2,3}$;~$^1\Delta$),
 $[...]\sigma_1^2\sigma_2^1\pi^1$~($^3\Pi_{0^-,0^+,1,2}$;~$^1\Pi$)
 were calculated. Here the $\sigma_1$ orbital is mainly formed by
 the $2p_z$ orbital of F with admixture of the $6p_z$ and $6s$ orbitals of Hf,
 $\sigma_2$ is mainly the $6s$ orbital of Hf with admixture of the
 $6p_z$ orbital of Hf, $\delta$ and $\pi$ are mainly the $5d$ orbitals of Hf.

 To obtain the spectroscopic parameters, six points listed in table \ref{dip}
 and a point at 100 a.u.\ on the HfF$^+$ potential curves were calculated for
 ten lowest-lying states in ten-electron calculations and for four states in
 20-electron ones. The 20-electron calculation is substantially more time consuming;
 therefore the remaining six states were calculated for only one point, 3.4 a.u., 
 in the present study. Comparing the latter calculations with corresponding
 ten-electron ones, the core ($5s^2$ and $5p^6$ shells of Hf and $1s^2$ shell of F) 
 relaxation and correlation corrections to the $T_e$ values, called
 ``core corrections'' below and in \tref{spec}, were estimated.
%  Since only the electronic $^1\Sigma^+$ state is below the $^3\Delta_1$ one,
%  the radiative lifetime of the lowest vibrational levels of the latter can be
%  written (after neglecting the rotation) as
Only the electronic $^1\Sigma^+$ state is below the $^3\Delta_1$ one.
The radiative lifetime of the lowest vibrational levels of the latter 
with respect to its decay to the vibrational levels of the $^1\Sigma^+$ state 
can be written (here we are neglecting the rotational structure of the
considered states) as

\begin{equation}
   \tau_{{\rm el}(^3\Delta_1,v2)}^{-1} = \frac{4}{3c^3}
  \sum_{
 v1(\Delta E > 0)
  }\Delta E^3
  |<\Psi_{^3\Delta_1,v2}|{\bf d}|\Psi_{^1\Sigma^+,v1}>|^2,
  \label{life}
\end{equation}
 where $\Delta E = (E_{(^3\Delta_1,v2)}- E_{(^1\Sigma^+,v1)})$ is the
 difference in the energies of the electronic-vibrational states
 $\Psi_{^3\Delta_1,v2}$ and $\Psi_{^1\Sigma^+,v1}$; $d$ is the dipole moment
 operator.  In the adiabatic approximation

\begin{eqnarray}
 \nonumber
  |<\Psi_{^3\Delta_1,v2}|{\bf d}|\Psi_{^1\Sigma^+,v1}>|^2 = \\
|<{\cal X}_{^3\Delta_1,v2}(R)|D(R)_{^1\Sigma^+ \rightarrow ^3\Delta_1}|{\cal X}_{^1\Sigma^+,v1}(R)>|^2,
  \label{adi}
\end{eqnarray}
 where ${\cal X}_{^3\Delta_1,v2}(R)$ is the vibrational wave function of the
 given electronic state and $D(R)_{^1\Sigma^+ \rightarrow ^3\Delta_1}$ is the
 electronic transition dipole moment as a function of the internuclear
 distance.  The excited vibrational levels of the $^3\Delta_1$ state can also
 decay to the lower vibrational levels of the same, $^3\Delta_1$, electronic
 state. The radiative lifetime $\tau_{\rm vibr}$ of this process is
 determined by the equations similar to (\ref{life}) and (\ref{adi}) with
 replacing $^1\Sigma^+$ by $^3\Delta_1$.  $D(R)_{^3\Delta_1 \rightarrow
 ^3\Delta_1}$ here is the total molecule-frame electric dipole moments for
 $^3\Delta_1$ state of the cation, calculated with respect to the center of
 mass.
% In the present work we calculate $D(R)$ (see \tref{dip}) for six points
% using the electronic wave functions obtained in ten-electron calculations.
 In \tref{dip} we tabulated $D^{10e}(R)$ for six points using the electronic
 wave functions obtained in the ten-electron calculations.  In \tref{dip20}
 we listed $D^{20e}(R)$ for point 3.4 a.u.\ using the electronic wave
 functions obtained in the twenty-electron calculations. To calculate the
 radiative lifetimes in according to eqs.\ (\ref{life}) and (\ref{adi}) we
 first use the $D^{10e}(R)$ values and then multiply the obtained $\tau_{\rm
 el}$ and $\tau_{\rm vibr}$ on the corresponding correction factors
 $(D^{10e}(3.4)/D^{20e}(3.4))^2$.  The vibrational wave functions ${\cal
 X}_{^3\Delta_1,v2}(R)$, ${\cal X}_{^1\Sigma^+,v1}(R)$ and
 electronic-vibrational energies were evaluated on the basis of the
 20-electron SODCI calculations.

 The ``atomic core'' properties $A_{\parallel}$ and $E_{\rm eff}$ were
 calculated only for the $^3\Delta_1$ state at the point 3.4 a.u., which is
 close to the equilibrium distance (see \tref{spec}).  Before calculating the
 core properties the shapes of the four-component molecular spinors are
 restored in the inner core region after the two-component GRECP calculation
 of the molecule. For this purpose the nonvariational one-center restoration
 method \cite{Titov:85Dis, Titov:96, Titov:96b, Titov:99, Petrov:02,
 Petrov:05a} is applied.  The main contribution to the electric field,
 $\bm{E}$, in eq.~\ref{Wd} gives the nuclear charge of Hf. In the present
 paper the Hf nucleus is modeled by a uniform charge distribution within a
 sphere with radius $R_{\rm nucl} = 6.8 {\rm fm}$. Taking into accounting
 that the electric field from electrons leads to a small correction (about
 one per cent \cite{Lindroth:89}) and the dominant contribution is going from
 the inner core electrons, we consider here only direct contributions of
 electrostatic interaction from the closed shells with $n=1-4$ with the EDMs
 of other (valence and outer-core) electrons.

%=====================================================================
\section{Results and discussion.}

\begin{figure*}
\includegraphics[scale=1]{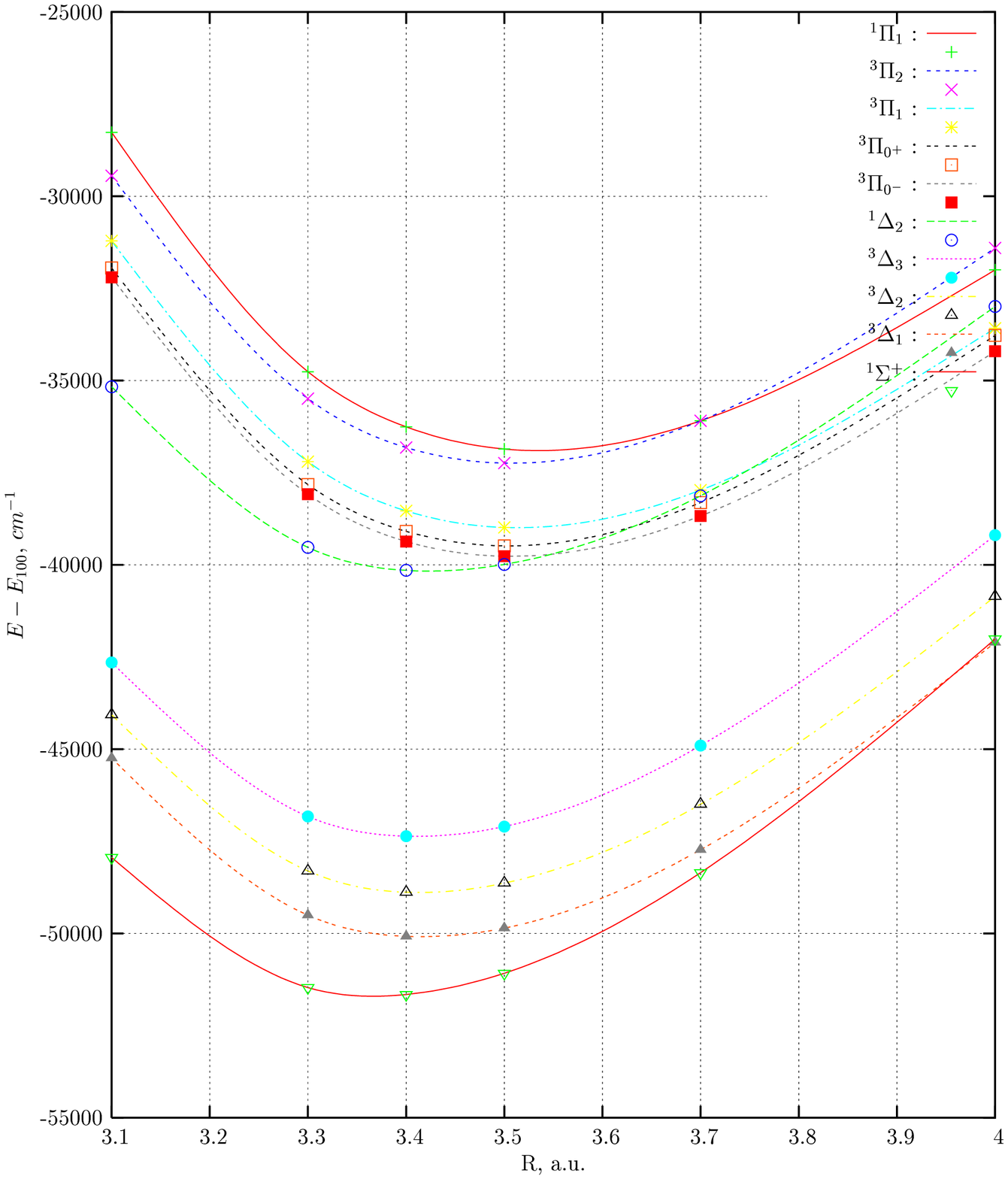}
\caption{\label{curves}  
Calculated potential energy curves of the HfF$^+$}
\end{figure*}

 The calculated potential energy curves of HfF$^+$ are shown in
%!!! do not translate with pdflatex!
 Fig.~\ref{curves}.
 The potential energy curves are taken from ten-electron calculations but
 shifted (on the energy axis) to fit the $D_e$ value obtained in the
 twenty-electron calculations and $T_e$ values obtained with twenty-electron
 correction at point $R=3.4 a.u.$ The results of calculations for HfF$^+$
 spectroscopic parameters are presented in \tref{spec}.  The first point to
 note is that the cation is deeply bound.  The second important result is
 that the $^1\Sigma^+$ state appears to be the ground one and the
 $^3\Delta_1$ state is the first excited one.
% So, additional study of schemes of populating the working state is
% required.
 In addition, the excitation energy from $^1\Sigma^+$ to $^3\Delta_1$ is
 increased from 866 to 1633 cm$^{-1}$ after including the $5s$ and $5p$
 shells of Hf and $1s$ shell of F into the relativistic correlation
 calculation.  Excitation energies from $^1\Sigma^+$ to other calculated
 low-lying states are also increased. Note also that the values obtained in
 ten-electron calculations for the lowest four states are in a good agreement
 with the purely 20-electron calculations when just the core correction
 described above is taken into account.  Accounting for correlation and
 relaxation of the $4f^{14}-$shell would also be desirable for better
 accuracy but it is too time consuming computationally. Though the $5s$ and
 $5p$ electrons are more strongly bound than the $4f$ electrons ($5s$ and
 $5p$ orbital energies about 2 and 1.2 times higher, respectively) one should
 rather expect that correlation and relaxation contibutions of the $4f$
 electrons will lead to a substantially smaller change of the spectroscopic
 properties of HfF$^+$.  There are several reasons for this: unlike the
 $5s,5p$ shells,  the $4f$ shell is localized in essentially different space
 region than the $5d$ and, particularly, $6s,6p$ ones (the average radii of
 the $5s$ and $5p$ shells are about twice larger than that of $4f$); the
 angular-type $4f{-}4f$ and $4f{-}6s,6p$ electron correlations are suppressed
 because there are no ``relatively low-lying'' $g$-states, whereas the
 lower-lying $4d$-states are completely occupied.  In \tref{hfd} the results
 of the Dirac-Fock calculations for the transition $5d^1 6s^1 \rightarrow
 5d^0 6s^2$ with $[Xe]4f^{14}5s^2 5p^6$ and $[Xe]4f^{14}$ frozen core are
 compared to the all-electron results. Two series of Dirac-Fock calculations
 are performed; in the first one the cores were taken being frozen from the
 $5d^16s^1$ state, and in the second one they were extracted from the
 $5d^06s^2$ state.  One can see from \tref{hfd} that the influence of $4f$
 relaxation is about one order of magnitude smaller than that of $5s5p$
 relaxation.
% Taking into account our remarks given above, we do not expect substantial
% change of the picture when correlations are taken into account.
 Similar picture is observed when correlations are taken into account.  The
 results of the all-electron Fock space relativistic coupled cluster (RCC)
 calculations for the transition from the terms of the $5d^1 6s^1$
 configuration to the  $5d^0 6s^2$ configuration are presented in \tref{rcc}.
 Three series of RCC calculations are performed;  In ten-electron calculation
 the $5s,5p$ shells are correlated, in  sixteen-electron calculation the $4f$
 shells are correlated, in twenty four-electron calculation the $4f,5s,5p$
 shells are correlated in addition to the 5d,6s,6p ones. So difference
 between the twenty four-electron and sixteen-electron calculations give us
 the correlation contribution of the $5s,5p$ shells, difference between the
 twenty four-electron and ten-electron calculations give us the contribution
 of the correlation of the $4f$ shell.  One can see from \tref{hfd} that the
 influence of $4f$ correlation is about one order of magnitude smaller than
 that of $5s5p$ correlation.  In the framework of the ``atom-in-a-molecule''
 model, the $5d^1 6s^1 \rightarrow  5d^0 6s^2$ excitation of the Hf$^{2+}$
 fragment gives the leading contribution to the $^1\Sigma^+ \rightarrow\
 ^3\Delta_1$ transition of HfF$^+$, and therefore we do not expect changes
 more than 100 cm$^{-1}$ in the energy for the above transition when $4f$
 correlation and relaxation are taken into account.  So we do not expect that
 the order of levels will be changed.
 The data of transition dipole moments from \tref{dip} and \tref{dip20} can
 be used to study schemes of populating the $^3\Delta_1$ state.

 The SO splittings of the $^3\Delta$ and $^3\Pi$ states are mainly due to the
 SO splitting of the $5d$ shell of Hf:

\begin{equation}
   {\bf H}_{ls}^{\rm so} = {\it a}\cdot({\bf l_{\rm 5d}}{\cdot}{\bf s_{\rm 5d}})\ ,
\label{hso}
\end{equation}
 The atomic Dirac-Fock calculation of Hf$^+$ gives
 $\veps_{5/2}{-}\veps_{3/2}=3173 {\rm ~cm}^{-1} \Rightarrow {\it a}=1269 {\rm
 ~cm}^{-1}$, where $\veps_{5/2}$ and $\veps_{3/2}$ are orbital energies of the
 $5d_{5/2}$ and $5d_{3/2}$ states of Hf$^+$. The SO interaction (\ref{hso})
 averaged over the $^3\Delta$ or $^3\Pi$ states is reduced to

\begin{equation}
   {\bf H}_{LS}^{\rm so} = {\cal A}\cdot({\bf L}{\cdot}{\bf S}),
\label{hso2}
\end{equation}
 where ${\cal A}= 1269/2 = 635 {\rm ~cm}^{-1}$, and $\bf{L}$ and $\bf{S}$ are the
 orbital momentum and spin moment of HfF$^+$. The SO interaction (\ref{hso2}) leads to
 splitting between the components of the $^3\Delta$ and $^3\Pi$ states on 1269 and
 635${\rm ~cm}^{-1}$, respectively. It is in good agreement with the
 splitting of the $^3\Delta$ state calculated in \tref{spec} but
%??? not
 with the $^3\Pi$ splitting because of the off-diagonal SO interaction (which
 is not described by expression (\ref{hso2})) with closely lying $^1\Delta$
 and $^1\Pi$ states.

 The calculated $A_{\parallel}$ and $E_{\rm eff}$ for the $^3\Delta_1$ state
 are presented in \tref{core}. In contrast to $A_{\parallel}[\rm{F}]$,
 $A_{\parallel}[\rm{Hf}]$ and $E_{\rm eff}[\rm{Hf}]$ are not seriously
 changed when the outer core electrons are included into the calculation.
 This behavior of $A_{\parallel}[\rm{F}]$ is explained by the fact that the
 shells of fluorine in the molecule can be considered with good accuracy as a
 closed-shell subsystem (thus having $A_{\parallel}[\rm{F}]\approx0$) in the
 $^3\Delta$ state, i.e.\ there is a large compensation of contributions from
 orbitals with different projections of total electronic momentum to
 $A_{\parallel}[\rm{F}]$.  Therefore, even a small perturbation
% i.e. explicit treatment 5s5p[Hf] & 1s[F]
  (influencing on the shells of F)
%  (of the shells contributing to $A_{\parallel}[\rm{F}]$)
 can seriously change the $A_{\parallel}[\rm{F}]$ value.  The calculated
 $E_{\rm eff}$ is large and comparable with the corresponding value for the
 $a(1)$ state of the PbO molecule \cite{Petrov:05a}.  Our $E_{\rm eff}$ value
 has opposite sign than that obtained by Meyer {\it et al.}\ \cite{Meyer:06a}
 in scalar-relativistic calculations and 1.34 times larger by absolute value.
 In \tref{lifetime} we are tabulating lifetimes $\tau_{\rm el}$ and
 $\tau_{\rm vibr}$ for the lowest vibrational levels of the $^3\Delta_1$ It
 is difficult to obtain this value accurately because of small absolute
 values of both transition energies and, particularly, of the transition
 dipole moments between those states, whereas the absolute errors are similar
 to those for other transitions.  Moreover, the transition energy and dipole
 moment calculated for the considered transition with rather large {\it
 relative errors}, are presented in equation (\ref{life}) as third and second
 powers, respectively, thus seriously increasing the relative error for the
 lifetimes of $^3\Delta_1$.  The correction factors
 $(D^{10e}(3.4)/D^{20e}(3.4))^2$ calculated from data of \tref{dip} and
 \tref{dip20} for $\tau_{\rm el}$ and $\tau_{\rm vibr}$ are 0.673 and 1.028,
 respectively.  
% At the moment we have no reliable tools to evaluate the uncertainty for this
% value.  As an illustration we note that  an error of $6.4\cdot10^{-3}$ a.u.\
% on the transition dipole moment, or 400 cm$^{-1}$ on the transition energy,
% which are rather reasonable for such a study, will change the calculated
% lifetime in two times for the given transition.
  Finally, it should be noted that the calculated energy for the 
  $^1\Sigma^+ \rightarrow {^3}\Pi_1$ transition is in a good agreement with 
  the pilot experimental datum by Cornell group \cite{Cornell:08}.
% Further increase of accuracy for this property can be better attained with
% improvement of the used {\sc sodci} code though study of similar systems,
% such as HfO, for which experimental data are available, can also be useful.

\begin{table}
\caption
%  {Electronic transition dipole moments (in $10^{-2}$ a.u.) for
%  $^3\Delta_1 \rightarrow ^1\Sigma^+$ %   in HfF$^+$}
{Ten-electron SODCI calculations of transition dipole moments 
and molecule-frame electric dipole moments for $^3\Delta_1$ state
calculated with respect to the center of mass. Axis $z$ is
directed from Hf to F.
All values in a.u.}
%\vspace{0.5 cm}
%\begin{ruledtabular}
\begin{tabular}{lcccccc}
\hline
\hline 
R               &    3.1   &     3.3    &  3.4     &    3.5    &     3.7   &   4.0  \\
 \vspace{-3 mm} \\
\hline
 \vspace{-3 mm} \\
%D(R)            &  1.60~  &    1.59~  &   1.60~ &  1.56~  & 1.50~ &     1.14~        \\
$D(R)_{^3\Delta_1 \rightarrow ^3\Delta_1}$               & -1.15~  &   -1.40~  &  -1.52~ & -1.64~  &-1.87~ & -2.20~  \\
$D(R)_{^1\Sigma^+ \rightarrow ^3\Delta_1} \cdot 10^{2} $ &  1.60~  &    1.59~  &   1.60~ &  1.56~  & 1.50~ &  1.14~  \\
$D(R)_{^1\Sigma^+ \rightarrow ^3\Pi_1} \cdot 10^{1} $    &  3.12~  &    2.94~  &   2.84~ &  2.77~  & 2.69~ &  2.70~  \\
$D(R)_{^1\Sigma^+ \rightarrow ^1\Pi_1} \cdot 10^{1} $    &  7.55~  &    6.52~  &   6.02~ &  5.59~  & 4.93~ &  4.11~  \\
$D(R)_{^3\Delta_1 \rightarrow ^3\Pi_1} \cdot 10^{2} $    &  3.78~  &    4.00~  &   4.13~ &  4.25~  & 4.51~ &  4.99~  \\
$D(R)_{^3\Delta_1 \rightarrow ^1\Pi_1} \cdot 10^{2} $    &  3.34~  &    3.90~  &   3.87~ &  3.93~  & 4.09~ &  4.50~  \\
$D(R)_{^3\Delta_1 \rightarrow ^3\Pi_{0+}} \cdot 10^{1} $ &  3.86~  &    3.30~  &   3.01~ &  2.80~  & 2.45~ &  2.13~  \\
\hline
\hline
\end{tabular}
\begin{flushleft}
\end{flushleft}
\label{dip}
\end{table}

\begin{table*}
\caption
{Transition and molecule-frame electric dipole moments (with respect
to the center of mass for R=3.4 a.u.) obtained in twenty-electron calculations. Axis
$z$ is directed from Hf to F.  All values in a.u.}
%\vspace{0.5 cm}
%\begin{ruledtabular}
\begin{tabular}{cccccccccc}
\hline
\hline 
                & $^1\Sigma^+$  & $^3\Delta_1$ & $^3\Delta_2$ & $^3\Delta_3$ & $^1\Delta_2$ & $^3\Pi_{0^+}$ &   $^3\Pi_1  $ & $^3\Pi_2$ &  $^1\Pi_1$    \\
 \vspace{-3 mm} \\
\hline
 \vspace{-3 mm} \\
%{ $^1\Sigma^+$ } & -1.20203     &  0.0195485827& 0 (6.651E-06)& 0 (-0.000219)& 0(-2.149E-05)&  -0.0777821   &  -0.35457281  & 0(-0.0001)& 0.532986998    \\
%{ $^3\Delta_1$ } &              &   -1.49835   &  0.013514591 & 0 (0.00518)  &  -0.0087410  &  0.330388978  &   0.02559     &  ?        &   0.00082299   \\
%{ $^3\Delta_2$ } &              &              & -1.483916    &  0.019299989 & 0.0317337    & 0 (3.733E-05) & 0.230772892   &  ???      &   0.2147366    \\
%{ $^3\Delta_3$ } &              &              &              & -1.49266898  &  -0.04693480 & 0 (-0.015016) & 0 (-0.001146) &  ?        & 0 (0.0043115)  \\
%{ $^1\Delta_2$ } &              &              &              &              & -1.29579     & 0(-0.0001515) &  0.172717514  &  ???      &  -0.214399762  \\
%{ $^3\Pi_{0^+}$} &              &              &              &              &              & -0.948399     & -0.0084163    &  ???      &   -0.00182371  \\
%{ $^3\Pi_1  $  } &              &              &              &              &              &               &  -1.048435    &  ?        &    -0.134798   \\
%{ $^3\Pi_2  $  } &              &              &              &              &              &               &               & -1.01587  &      ?         \\
%{ $^1\Pi_1  $  } &              &              &              &              &              &               &               &           &    -1.2374946  \\
{ $^1\Sigma^+$ } &   -1.20      & 1.95$\cdot10^{-2}$ & 0      & 0 & 0            & 7.77$\cdot10^{-2}$ & 3.55$\cdot10^{-1}$   & 0 &   5.33$\cdot10^{-1}$  \\
{ $^3\Delta_1$ } &              &   -1.50      &  1.35$\cdot10^{-2}$ & 0     & 8.74$\cdot10^{-3}$ & 3.30$\cdot10^{-1}$   &  2.56$\cdot10^{-2}$ & 8.42$\cdot10^{-3}$         &   8.23$\cdot10^{-4}$      \\
{ $^3\Delta_2$ } &              &              & -1.48        &   1.93$\cdot10^{-2}$ & 3.17$\cdot10^{-2}$    & 0             & 2.31$\cdot10^{-1}$         & -2.10$\cdot10^{-2}$   &  2.15$\cdot10^{-1}$  \\
{ $^3\Delta_3$ } &              &              &              & -1.49        & 4.69$\cdot10^{-2}$ & 0      & 0             & 0.295     & 0              \\
{ $^1\Delta_2$ } &              &              &              &              & -1.30        & 0             &  1.73$\cdot10^{-1}$       & 4.17$\cdot10^{-2}$         &  2.14$\cdot10^{-2}$        \\
{ $^3\Pi_{0^+}$} &              &              &              &              &              & -9.48$\cdot10^{-1}$        & 8.42$\cdot10^{-3}$             &  0        &  1.82$\cdot10^{-3}$     \\
{ $^3\Pi_1  $  } &              &              &              &              &              &               &  -1.05        & 2.95$\cdot10^{-2}$    &  1.35$\cdot10^{-1}$      \\
{ $^3\Pi_2  $  } &              &              &              &              &              &               &               & -1.02     & -5.79$\cdot10^{-2}$      \\
{ $^1\Pi_1  $  } &              &              &              &              &              &               &               &           &    -1.24       \\
\hline
\hline
\end{tabular}
\label{dip20}
\end{table*}

%--------------------------------------------------------------
%\squeezetable
\begin{table}
\caption
  {Calculated spectroscopic parameters for HfF$^+$} 
%\vspace{0.5 cm}
%\begin{ruledtabular}
\begin{tabular}{ccrrcc}
\\
\hline
\hline
 \vspace{-3 mm} \\
  State  & $R_e$ $\AA$~~ & $T_e$ $cm^{-1}$ & ~~$T_e$ with core & ~~$w_e$ $cm^{-1}$ & $D_e$ $cm^{-1}$ \\
       &               &                 & correction$^a$ &                   &                 \\
\multicolumn{6}{c}{}
 \vspace{-3 mm} \\
\hline
 \vspace{-3 mm} \\
\multicolumn{5}{c}{\bf 10-electron calculation}\\
 \vspace{-3 mm} \\
\hline
 \vspace{-3 mm} \\
     { $^1\Sigma^+$ } & 1.784    &     0   &     0  &  751  & 51107 \\
     { $^3\Delta_1$ } & 1.810    &   866   &  1599  &  718  & \\
     { $^3\Delta_2$ } & 1.809    &  1821   &  2807  &  719  & \\
     { $^3\Delta_3$ } & 1.807    &  3201   &  4324  &  721  & \\
     { $^1\Delta_2$ } & 1.814    &  9246   & 11519  &  696  & \\
     { $^3\Pi_{0^-}$} & 1.856    &  9466   & 11910  &  689  & \\
     { $^3\Pi_{0^+}$} & 1.854    &  9753   & 12196  &  699  & \\
     { $^3\Pi_1  $  } & 1.860    & 10190   & 12686  &  687  & \\
     { $^3\Pi_2  $  } & 1.856    & 11898   & 14438  &  703  & \\
     { $^1\Pi_1  $  } & 1.870    & 12642   & 14784  &  679  & \\
 \vspace{-3 mm} \\
\hline
 \vspace{-3 mm} \\
\multicolumn{6}{c}{\bf 20-electron calculation}\\
 \vspace{-3 mm} \\
\hline
 \vspace{-3 mm} \\
     { $^1\Sigma^+$ } & 1.781    &     0   &        & 790  & 51685 \\
     { $^3\Delta_1$ } & 1.806    &  1633   &        & 746  & \\
     { $^3\Delta_2$ } & 1.805    &  2828   &        & 748  & \\
     { $^3\Delta_3$ } & 1.804    &  4273   &        & 749  & \\
 \vspace{-3 mm} \\
\hline
\hline

\end{tabular}
% \noindent
\begin{flushleft}
$^{\rm a}$ See section ``Methods and calculations'' for details.\\
\end{flushleft}
%\end{ruledtabular}
\label{spec}
\end{table}
%--------------------------------------------------------------

%----------------------------------------------------------------------
\begin{table}
\caption{ 
   Transition energies (in cm$^{-1}$) for the $5d^16s^1 \rightarrow  5d^06s^2$ 
   obtained in Dirac-Fock calculations of the Hf$^{2+}$ ion.
   }
\begin{tabular}{cccc}
\\
\hline
\hline
Frozen      &   $[Xe]4f^{14}5s^25p^6$      &   $[Xe]4f^{14}$    & all electron \\
core        &                              &                    &  included    \\
 \vspace{-3 mm} \\
\hline
 \vspace{-3 mm} \\
Frozen from &          13~322              &        12~621      &    12~439     \\ 
 $5d^16s^1$ &                              &                    &               \\
Frozen from &          11~499              &        12~264      &    12~439     \\
$5d^06s^2$  &                              &                    &               \\
\hline
\hline
\end{tabular}
\label{hfd}
\end{table}

\begin{table*}
\caption{ 
   Transition energies (in cm$^{-1}$) from the terms of the $5d^16s^1$
   configuration to the  $5d^06s^2$ configuration 
   obtained in RCC calculations of the Hf$^{2+}$ ion.
   }
\begin{tabular}{cccccc}
\\
\hline
\hline
Transition  &   ten-electron      &   sixteen-electron    &   twenty four-electron & conribution  & conribution \\
 from       &     calculation     &     calculation       &      calculation        & from $5s5p$  & from $4f$  \\
 \vspace{-3 mm} \\
\hline
 \vspace{-3 mm} \\
$6s_{1/2}5d_{3/2}(J=1)$ & 9233    &      10899            &         9182            &    -1717     &    -51     \\
$6s_{1/2}5d_{3/2}(J=2)$ & 8935    &      11353            &         8817            &    -2536     &   -118     \\
$6s_{1/2}5d_{5/2}(J=2)$ & 6251    &      8763             &         6036            &    -2727     &   -215     \\
$6s_{1/2}5d_{5/2}(J=3)$ & 5168    &      6991             &         4973            &    -2018     &   -195     \\
\hline
\hline
\end{tabular}
\label{rcc}
\end{table*}

%--------------------------------------------------------------
\begin{table}
\caption{
   Calculated parameters $A_{\parallel}[\rm{Hf}]$ and
   $A_{\parallel}[\rm{F}]$ (in MHz) and $E_{\rm eff}$ (in $10^{24}{\rm
   Hz}/(e\cdot{\rm cm})$) for the $^3\Delta_1$ state of
   $^{177}$Hf$^{19}$F$^+$ at internuclear 
   distance of 3.4 a.u.
}

\begin{tabular}{ccc}
\\
\hline
\hline
 \vspace{-2 mm} \\
~~$A_{\parallel}[\rm{Hf}]$~~  & ~~$A_{\parallel}[\rm{F}]$~~~ & ~~~$E_{\rm eff}$~~ \\
 \vspace{-2 mm} \\
\hline
%\\
 \vspace{-3 mm} \\
\multicolumn{3}{c}{\bf 10-electron calculation } \\
 \vspace{-3 mm} \\
\hline
 \vspace{-2 mm} \\
-1250 &  -33.9   &  5.89  \\
 \vspace{-2 mm} \\
\hline
 \vspace{-3 mm} \\
\multicolumn{3}{c}{\bf 20-electron calculation } \\
 \vspace{-3 mm} \\
\hline
 \vspace{-2 mm} \\
 -1239     &   -58.1       &   5.84 \\
 \vspace{-2 mm} \\
\hline
\hline
\end{tabular}
\label{core}
\end{table}

\begin{table}
\caption{
   Calculated vibrational energy levels (in cm$^{-1}$) of the $^1\Sigma^+$ 
   and $^3\Delta_1$ states and radiational lifetimes (in seconds) of the
   $^3\Delta_1$ state.
}

\begin{tabular}{ccccc}
\\
\hline
\hline
 \vspace{-2 mm} \\
{\it v} &  $^1\Sigma^+$ &  \multicolumn{3}{c}{$^3\Delta_1$} \\
 \vspace{-2 mm} \\
\hline
%\\
   &    \multicolumn{2}{c}{$E_v$}  & $\tau_{\rm el}$  &  $\tau_{\rm vibr}$ \\ 
 0~ &      416~    &     368~~~     &   0.389~         &      -             \\
 1~ &     1234~    &    1117~~~     &   0.365~         &    0.184           \\
 2~ &     2038~    &    1878~~~     &   0.343~         &    0.088           \\
 3~ &     2821~    &    2631~~~     &   0.316~         &    0.059           \\
 4~ &     3586~    &    3373~~~     &   0.290~         &    0.046           \\

\hline
\hline
\end{tabular}
\label{lifetime}
\end{table}

%--------------------------------------------------------------

%\paragraph*{Conclusion.}
%pra
%\paragraph*{Acknowledgments.}
%end pra
%
\section{Acknowledgments.}

 N.M.\ and A.T.\ are grateful to RFBR for grant No.\ 06--03--33060, the
 partial support from RFBR grant No.\ 07--03--01139 is also acknowledged.
%ap
% A.P.\ is grateful for grants No.\ 29-04/32 from St.\ Petersburg Committee on
% Science and Higher Education.
A.P.\ is also supported by grant of the Governor of Leningrad district.
%end ap 
 The authors thank E.Cornell for stimulating
 this work and for many useful discussions.

\bibliographystyle{./bib/apsrev}

%\bibliography{bib/*}

%\bibliography{bib/JournAbbr,bib/Titov,bib/TitovLib,bib/Kaldor,bib/Isaev}
\bibliography{bib/JournAbbr,bib/Titov,bib/TitovLib,bib/TitovAbs,bib/Kaldor,bib/Isaev,bib/PetrovLib}

\end{document}